\documentclass[aps,twocolumn,prl,superscriptaddress,showpacs,preprintnumbers,amsmath,amssymb]{revtex4}

\usepackage{graphicx}
\usepackage{color}

\begin{document}

\title{Toward
High Temperature Quasi-two-dimensional Superconductivity}

\author{V.~V.~Kabanov}
\affiliation{Zavoisky Physical-Technical Institute, FIC KazanSC of
RAS, 420029 Kazan, Russia} \affiliation{Department for Complex
Matter, Jozef Stefan Institute, 1000 Ljubljana, Slovenia}

\author{I.\,I.\,Piyanzina}
\affiliation{Zavoisky Physical-Technical Institute, FIC KazanSC of
RAS, 420029 Kazan, Russia} \affiliation{Institute of Physics, Kazan
Federal University, 420008 Kazan, Russia}

\author{D.\,A.\,Tayurskii}
\affiliation{Institute of Physics, Kazan Federal University, 420008
Kazan, Russia}

\author{R.~F.~Mamin}
\affiliation{Zavoisky Physical-Technical Institute, FIC KazanSC of
RAS, 420029 Kazan, Russia} \affiliation{Institute of Physics, Kazan
Federal University, 420008 Kazan, Russia}

\date{\today}

\begin{abstract}
The demonstration of a quasi-two-dimensional electron gas (2DEG) and
superconducting properties in LaAlO$_3$/SrTiO$_3$ heterostructures
has stimulated intense research activity in recent ten years. The
2DEG has unique properties that are promising for applications in
all-oxide electronic devices. The superconductivity in such
heterostructures has been observed below 300 mK. For
superconductivity applications it is desirable to have more wide
temperature of the existence range and the ability to control
superconductivity properties by external stimulus. Based on
first-principles calculations and theoretical consideration we show
that all-oxide heterostructures incorporating ferroelectric
constituent, such as BaTiO$_3$/La$_2$CuO$_4$, allow creating 2DEG.
We predict a possibility of a high temperature guasi-two-dimensional
superconductivity state. This state could be switchable between
superconducting and conducting states by ferroelectric polarization
reversal. We also discuss that such structures must be more simple
for preparation. The proposed concept of ferroelectrically
controlled interface superconductivity offers the possibility to
design novel electronic devices.
\end{abstract}

\maketitle

The creation of quasi-two-dimensional superconducting states at the
interface and the ability to control such states by magnetic and
electric fields is impossible without the use of new materials and
without the development of new design interfaces. Unique properties
of functional materials are achieved due to the effects associated
with the complex composition of the interface structure. Such new
materials include oxide heterointerfaces between two nonconducting
oxides in which, owing to strong electronic correlations, unique
transport properties are observed. A quasi-two-dimensional electron
system (2DEG) has been discovered~\cite{S1} at the interface between
two oxide insulators, LaAlO$_3$ (LAO) and SrTiO$_3$ (STO) by Ohtomo
and Hwang~\cite{S1},
 and it has attracted significant
attention~\cite{S1,S2,S3,S4,S5,S22,S7,S8,S9} due to a wide range of
other physical phenomena observed in LAO/STO. High carrier mobility
and high electron density of the 2DEG were
demonstrated~\cite{S1,S3,S7,S8,S9}, which makes it promising for
applications in all-oxide field-effect devices. Subsequently, the
coexistence of a two-dimensional electron superconductivity and
ferromagnetism was discovered in this system~\cite{S3,S4}. It was
found~\cite{S3} that the system passes into the superconducting
state below 300 mK. The density of the charge carriers in such a
heterostructure reaches the value of 3$\cdot $10$ ^{13}$\,cm$^{-2}$.

The most common mechanism for describing these phenomena is the
polarization catastrophe model~\cite{S1,S12}. The polar
discontinuity at the interface leads to the divergence of the
electrostatic potential. Along the [001] direction, LAO can be
considered as an alternation of the differently charged layers of
(LaO)$^{+1}$ and (AlO$_2)^{-1}$. As it was shown experimentally, in
the heterostructure with the TiO$_2$ interface layer the electric
potential along the [001] direction appears due to the polarity
disruption at the interface. Thus, the atomically flat quality of
the interface between two components is utterly necessary since the
effect is related to the strictly defined sequence of layers inside
each slab. A transition to the superconducting state is observed at
very low temperatures. That is why it is essential to develop
technical approaches to create quasi-two-dimensional
superconductivity at higher temperatures and it is also important to
study the processes of switching superconductivity.

We present the results of the $ \textit{ab-initio} $ calculation of
the structural and electronic properties  of the heterostructure
consisting of ferroelectric material and parent compound of high
temperature superconductor (PCHTSC) of  BaTiO$_3$/La$_2$CuO$_4$
(BTO/LCO) heterostructure. We consider a possibility of a high
temperature quasi-two-dimensional superconductivity (HT2DSC) state
appearance in that heterostructure. We discuss the
Kosterlitz-Thouless critical temperature $T_{KT}$ for the transition
to the superconducting state due to preformed Cooper pairs. We also
show that such structures must be more simple for preparation.

For densities of states calculations and structural optimization
 we have
used density functional theory (DFT)~\cite{S14}. Exchange and
correlational effects were accounted by generalized gradient
approximation (GGA)~\cite{S17}. Kohn-Sham equations were solved
using the plane-wave basis set (PAW)~\cite{S18}, realized within the
VASP code~\cite{S15}, which is a part of the
MedeA\textsuperscript{\textregistered} software of Materials Design
\cite{S18}. The cut-off energy was chosen to be 400\,eV. The force
tolerance was 0.05~eV/\AA\ and the energy tolerance for the
self-consistency loop was $ 10^{-5} $~eV. The Brillouin zones were
sampled including $ 5 \times 5 \times 1 $ $ {\bf k} $-points. Since
there is a strong correlation between $ d $ and $ f $-electrons in
our system the GGA+$ U $ correction were included to our
computational scheme~\cite{S16}. The $ U $ parameter was added to
La\,4$ f $, Ti\,3$ d $ and Cu\,4$ d $ orbitals ($ U $=8\,eV, 2\,eV
and 10\,eV, respectively). The choice of $ U $ for Ti and La values
was based on our previous research~\cite{piyanzina}. The choice of $
U $ for Cu was based on comparison with LSDA+$U$ calculations and
experimental data for band gap and Cu local magnetic moment (Table 2
from~\cite{LCO_data} and \cite{svane}).

\begin{figure}[h!]
    \centering{\includegraphics[angle=-90,width=8cm]{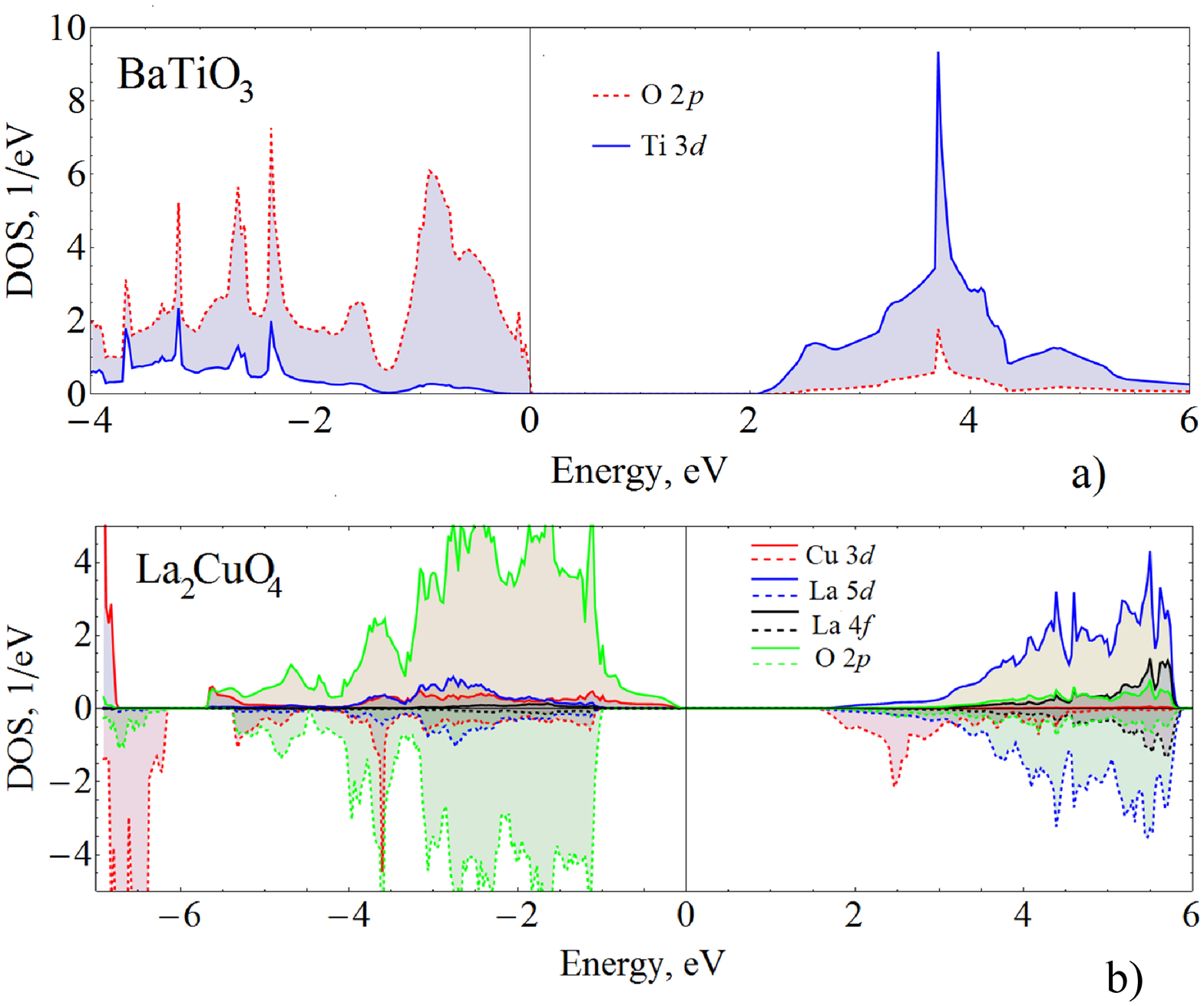}}
       \caption{The density of states of the bulk BaTiO$_3$ in the tetragonal phase (a) and La$_2$CuO$_4$  in the orthorhombic phase (b)}
    \label{bulk_doses}
\end{figure}

We study the heterostructure components separately. The structure
optimization has been performed  for the bulk BTO with quasi-cubic
tetragonal structure and for the bulk LCO with orthorhombic
structure. Cell parameters obtained after optimization were
$a$=$b$=4.00\,\AA, $c$=4.02\,\AA\ for BTO and $a$=5.42\,\AA,
$b$=5.42\,\AA, $c$=13.23\,\AA\  for LCO (experimental values:
$a$=$b$=3.999\,\AA, $c$=4.022\,\AA\ and $a$=5.331\,\AA,
$b$=5.339\,\AA, $c$=13.150\,\AA ,\  respectively~\cite{exp_BTO}). In
Fig.\,\ref{bulk_doses}\,(a) the density of states spectrum of the
bulk BTO material is presented. The obtained band gap is somewhat
lower than experimental value of 3.2\,eV~\cite{exp_BTO2}. In order
to determine the electronic properties of the studied structure the
density of states (DOS) spectrum has been calculated taking into
account magnetic properties of LCO. Fig.\,1b shows the atom-resolved
DOS for La, Cu and O. It is seen that the band gap of bulk LCO is
approximately 1.55 eV.

\begin{figure}
    \centering{\includegraphics[angle=-90,width=7cm]{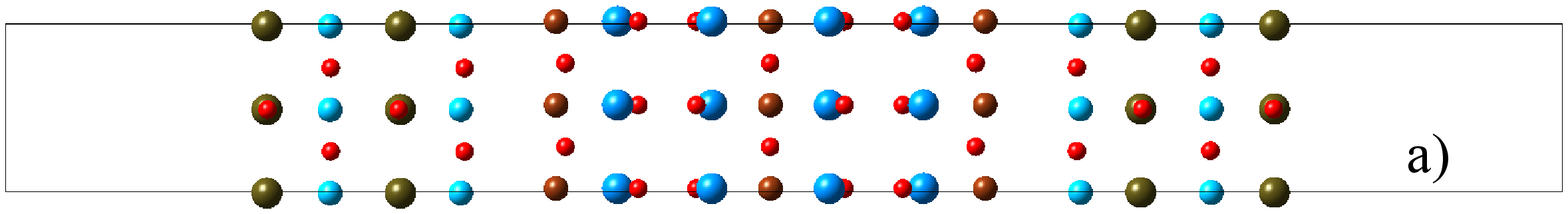}}
        {\includegraphics[angle=-90,width=8cm]{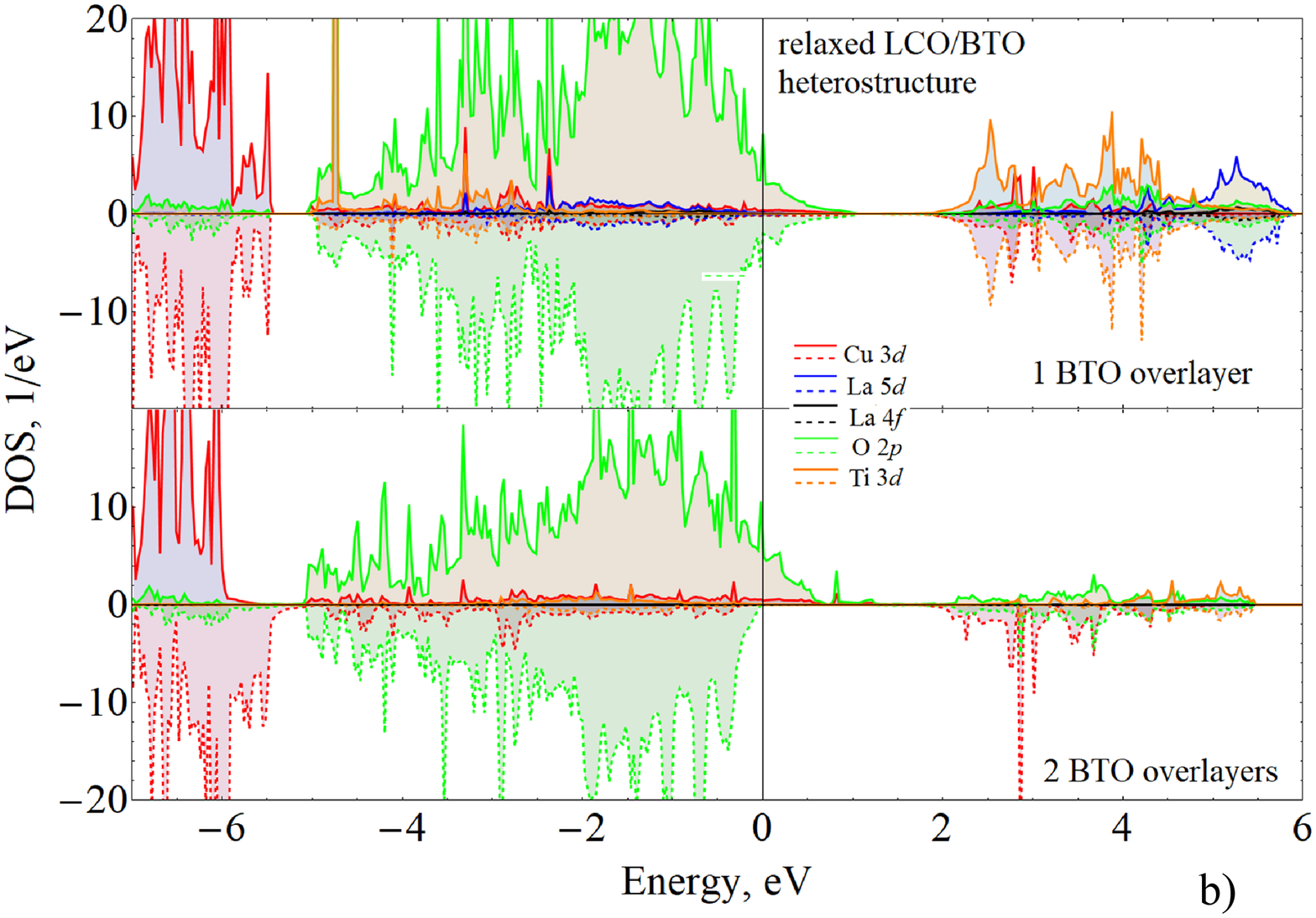}}
        \caption{ The unit cell of the heterostructure BaTiO$_3$/La$_2$CuO$_4$  (a) and
    their density of states (b)}
    \label{structure}
\end{figure}

In Fig.\,2a half of the unit cell of the studied system BTO/LCO with
BaO interface layer is presented whereas the second part is a mirror
copy with respect to the central LaO layer.
%
%
For modeling
the heterostructure the LCO central slab  was enlarged by a factor
of 1.5 and bounded  by a varying number of BaTiO$_3$ layers with
interface BaO or TiO$_2$ layers on both sides. Such a unit cell
guarantees the absence of the dipole moment and additional polarity
which might arise due to non-symmetric structure. In order to avoid
interaction of the surfaces and slabs with their periodic images, a
20~\AA\ vacuum region was added. In plane $a$ and $b$ cell
parameters were fixed, whereas atom positions were allowed to relax
during the optimization procedure.

After the optimization of both interface types  (first corresponds
to BaO interface layer, second -- to the TiO$ _{2} $ layer) it was
found that in the first case, the total energy of the system is
lower, meaning that the structure is more stable. That is why all
further reasoning will be presented for the most stable
configuration. It is seen from Fig.~2a that in the near-surface
TiO$_{2} $ layer the Ti atoms move out of the oxygen planes by  $a
\approx $0.15\,\AA. That leads to a dipole moment induction towards
the interface. Calculations involving higher number of the BTO
layers are required to get a full picture of structural distortions,
what will be done in our further publications.

In order to determine the electronic properties of the studied
structure the density of states (DOS) spectrum has been calculated
taking into account magnetic features of LCO. Fig.\,2b show the
atom-resolved DOS for Cu, La and Ti for 1 and 2 BTO overlayers. It
is seen that already with one BTO layer the band gap is closed. Cu
atomic orbitals cross the Fermi-level. Besides, the total magnetic
moment induction takes place which is mainly corresponds to Cu atoms
forms.

Let us analyze the results and perform some estimates of the
parameters of the arising state. We could  estimate
 the width of the area with metallic conductivity
 as 0.7-3 nm. Then for a two dimensional density
$n_s$ we got the value: $n_s\approx 10^{14}-4\cdot 10^{14}\
$cm$^{-2}$. Thus we expect that the system will become
superconducting. Taking into account that the thickness of the of
the conducting layer is small the superconducting properties are
governed by the Kosterlitz-Thouless transition. The temperature of
the transition is defined as
\begin{equation}
T_{KB}={A\hbar^2n_s\over{4 k_B m^*}},
\end{equation}
where $m^*$ is the effective mass of the current carriers, $A$ is a
coefficient ($A\approx $ 0.9 for two dimensional case)~\cite{S30}.
Taking into account that 2D density is $n_s= 10^{14}\,$cm$^{-2}$ and
assuming that $m^*=3m_e$~\cite{S31}, where $m_e$ is the free
electron mass, we obtain $T_{BK}\approx 70 K$. Note, that this
temperature is higher than the mean field critical temperature of
the bulk LCO with optimal doping($T_c\approx40K$). It means that the critical
temperature of the interface will be determined by the mean field
critical temperature of LCO with optimal doping.

Let us consider the effect of application of an external magnetic
field to SC leads. First of all note that the effective penetration
depth will be enhanced $\lambda_{eff}=\lambda^2/d$ (d is the
effective thickness of the interface layer). Therefore the lower
critical field will be strongly reduced. Depending on the intensity
of the magnetic field H, it either penetrates or does not penetrate
into the SC. The field does not penetrate into the SC at fields
below H$_{c1}$. Above H$_{c1}$, the field begins to penetrate into
the SC in the form of vortices, while the volume around the vortices
remains in the SC state. Fields above H$_{c2}$ completely penetrate
the sample, the magnetic field in the sample becomes uniform, and
the SC completely collapses. In the case of a thin SC film or a thin
SC layer, there will be the feature, primarily associated with a
demagnetizing factor. When the field is applied perpendicularly to
the SC layer the field at the film boundary increases greatly by a
factor of K (K$\approx 2-10\cdot 10^6$ for a 1-5 nm think
film~\cite{brandt}) due to a large demagnetizing factor. As a
result, when the field at the boundary H$_1$ = K H exceeds Hc1,
magnetic field vortices begin to penetrate the sample. Thus, the
value of the external magnetic field at which the field begins to
penetrate into SC is H$_{c1eff}$ = H$_{c1}$/K, where Hc1 is the
value for the bulk sample (in fact H$_{c1}$ for films is smaller
than that for bulk due to a transmission coefficient being bigger
for films. But the coefficient K is so huge, that H$_{c1eff}$ is
very small in any case). Thus, the value of H$_{c1eff}$ for a thin
SC layer turns out to be H$_{c1eff}\approx 0$ ($\ll$1 G). The value
of H$_{c2}$ for a thin SC layer is the same as that for a bulk. The
value of Hc2 for a thin SC layer is still large (for example,
H$_{c2}\approx $ 390000 G for La$_{0.85}$Sr$_{0.15}$MnO$_3$).
Therefore, it is necessary to expect the appearance of flux flow
resistivity in our heterostructure in any arbitrarily small fields,
due to the scattering of the superconducting current by flowing
vortices.

\begin{figure}
    \centering{\includegraphics[angle=-90,width=8cm]{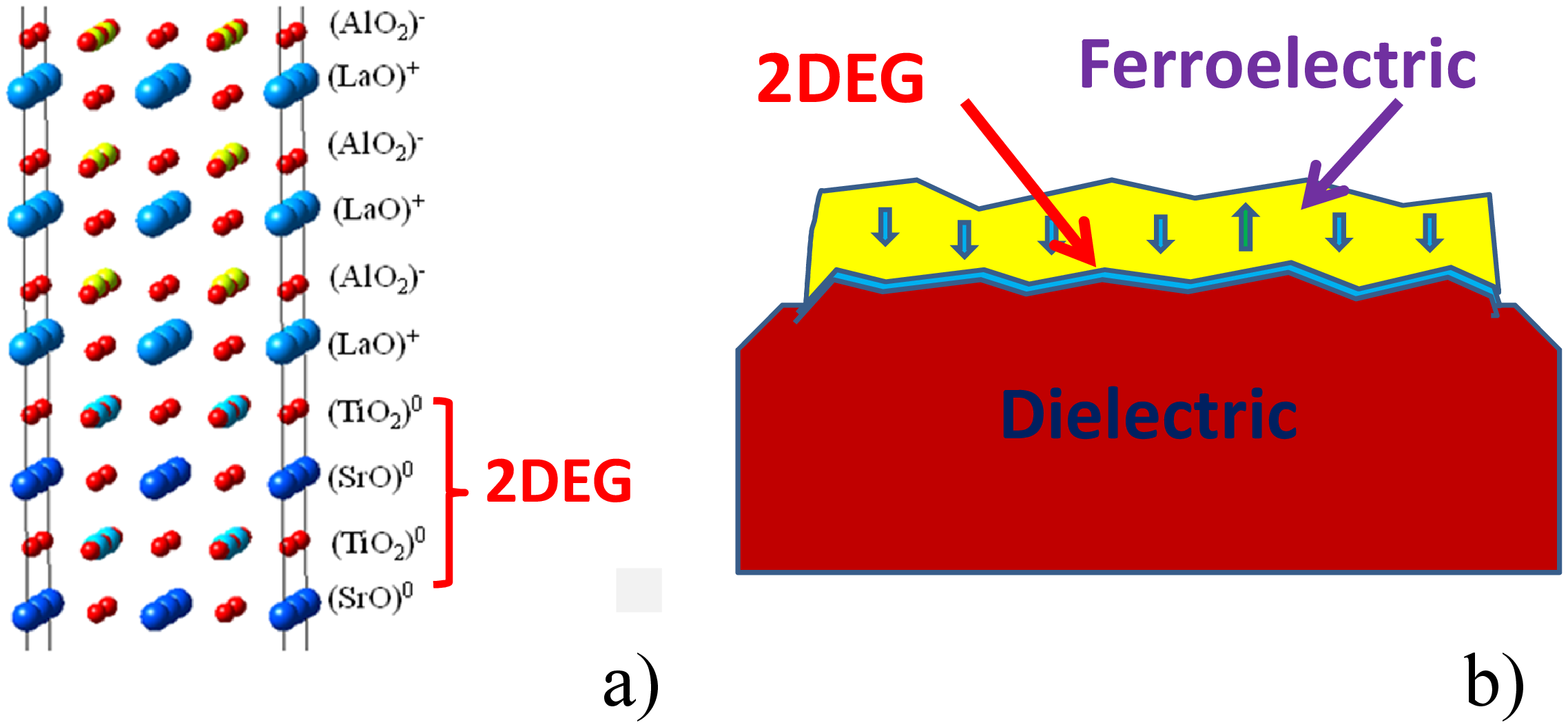}}
        \caption{ The heterostructure LAO/STO, interface area (a) and
        schematic illustration of the ferroelectric/dielectric with non-flat interface boundary
        (b)}
    \label{structure}
\end{figure}

It is supposed that the appearance of interface conductivity is
related to the structural and, consequently, electronic
reconstructions. The layers (BaO)$^0$ and (TiO$_2$)$^0$ are
"electrically neutral" in the simplest ionic limit, but there is a
ferroelectric polarization due to the Ti$^{+4}$ atoms displacements
out of octahedron center in the BTO. The direction of such a
polarization can be switched by an external electric field. That is
impossible to  do in the case of LAO slab, because an external
influence cannot change the sequence of (LaO)$ ^{1+} $ and (AlO$
_{2})^{1-}$ layers. Moreover it is very important that in this case
there is no need to make a very high-quality interface boundary,
because the polarization arises in the volume of the ferroelectric.
This differs from the case of LAO where, for the appearance of
polarization on the interface, it is necessary to obey strictly the
sequence of (LaO)$^{+1}$ and (AlO$_2)^{-1}$ layers. Besides, the
BTO/LCO system attracts the interest because it contains
antiferromagnetic insulator La$_2$CuO$_4$, which can be transferred
to conducting and superconducting state by increasing the
concentration of the free carriers~\cite{Mul,LaSr}. That was
realized by doping of the LCO by Sr or Ca~\cite{Mul,LaSr}. It can be
further expected that increasing the free charge carriers can lead
to the 2D superconductivity in a system with 2DEG. Therefore, there
is an opportunity to switch both conductivity and superconductivity
by an electric field in the heterostructures similar to BTO/LCO.

Recently, we have experimentally observed the occurrence of a
quasi-two-dimensional state with a metallic character of the
temperature behavior of the conductivity at the boundary of the
ferroelectric Ba$_{0.8}$Sr$_{0.2}$TiO$_3$ and the antiferromagnet
LaMnO$_3$~\cite{S62}. A distinctive feature of these experiments was
that the 2DEG state occurs under conditions where the boundary
between regions with different components at the interface in the
heterostructure is not ideally flat. In LAO/STO 2DEG formation
requires the atomically flat boundary and, in addition, a certain
sequence of layers of LaO and AlO$_2$. Earlier it was
predicted~\cite{S22,SS7} that 2DEG can arise when a ferroelectric
film is applied to an insulator. In this paper, we present an idea
that two-dimensional superconductivity at relatively high
temperatures can arise when a ferroelectric film is grown at the an
insulator substrate which could become a superconductor  after
doping.
%
%

The work ~\cite{S62} was the first research in the low interface
quality systems. Thus the proposed heterostructure seems to be
exceptionally crucial. Firstly, we investigate the emergence of a
quasi-two-dimensional electron gas at a new type of
BaTiO$_3$/La$_2$CuO$_4$ heterostructure interface in which there is
neither LaAlO$_3$ nor SrTiO$_3$. Secondly, PSHTSC can transform into
a superconductor state with an increase in the number of carriers,
so new physics and new possibilities appear here. And the most
importantly, thirdly, the use of a ferroelectric  makes it possible
to avoid the necessity of an extremely high quality of the
interface, because the polarization in BaTiO$_3$ allow the
polarization catastrophe to occur at any interface quality. The
possibility of a quasi-two-dimensional electron gas creation under
less stringent conditions for the interface quality of the
heterostructure and the realization of multifunctional conductivity
switching regimes and magnetization can substantially increase the
possibility of utilization these systems in the mass production of
technical devices based on such heterostructures.

Earlier, in the work of Bozovic group the giant proximity effect in
cuprate superconductors had been observed~\cite{Bozovic}. Later, in
Goldmann's work~\cite{S61} the switching-off of the
superconductivity when an electric field is applied to a sample with
superconducting properties through an ionic conductor  was
considered. In our investigation, we propose the creation of a
superconducting state due to the proximity effect with a
ferroelectric, therefore turning it off and turning it on with
electric or magnetic fields, by changing the direction of
polarization in the ferroelectric film, or by changing the magnetic
properties of the substrate, respectively. We also emphasize that
when ferroelectric films are used in heterostructures, the less
stringent requirements are imposed on the quality of the emerging
interfaces. Therefore, the suggested interfaces with ferroelectric
films are a completely new approach to the creation of 2DEG, which
we have already tested~\cite{S62}. And the idea of creating on the
basis of such 2DEG states of two-dimensional superconductivity at
the interface with LCO is even more unique proposal, which has no
analogues in the world.

In conclusion, in our paper we have presented the calculations of
the structural and electronic properties of the ferroelectric/PCHTSC
(BaTiO$_3$/La$_2$CuO$_4$) heterostructure. After the epitaxial
ferroelectric BaTiO$_3$ film was deposited on the LCO sample using,
for example, the magnetron sputtering technique we can get HT2DSC in
the interphase.

The authors from Kazan Federal University acknowledge partial
support by the Russian Government Program of Competitive Growth of
Kazan Federal University. The reported study was supported by the
Supercomputing Center of Lomonosov Moscow State University. R.F.M.
acknowledges financial support from Slovenian Research Agency,
Project BI-RU/16-18-021.


\begin{thebibliography}{24}

\bibitem{S1}
A.~Ohtomo and H.~Y.~Hwang, Nature {\bf 427}, 423 (2004), {\bf 5},
204 (2006).

\bibitem{S2}
S.~Thiel, G.~Hammerl, A.~Schmehl, C.~W.~Schneider, and J.~Mannhart,
Science {\bf 313}, 1942 (2006).

\bibitem{S3}
N.~Reyren, S.~Thiel, A.~D.~Caviglia, L.~Fitting~Kourkoutis,
G.~Hammerl, C.~Richter, C.~W.~Schneider, T.~Kopp,
A.-S.~R{\"u}etschi, D.~Jaccard,M.~Gabay, D.~A.~Muller,
J.-M.~Triscone, and J.~Mannhart, Science {\bf 317}, 1196 (2007).

\bibitem{S4}
A.~Brinkman, M.~Huijben, M.~Van Zalk, J.~Huijben, U.~Zeitler,
J.~C.~Maan, W.~G.~van der Wiel, G.~Rijnders, D.~H.~A.~Blank, and
H.~Hilgenkamp, Nature Mater.~{\bf 6}, 493 (2007).

\bibitem{S5}
A.~Kalabukhov, R.~Gunnarsson, J.~B{\"o}rjesson, E.~Olsson,
T.~Claeson, and D.~Winkler, Phys.~Rev.~B {\bf 75}, 121404 (2007).

\bibitem{S22}
M.\, K.~Niranjan, Y.~Wang, S.\,S.~Jaswal, and E.\,Y.~Tsymbal,
Phys.~Rev.~Lett. {\bf 103}, 016804 (2009).

\bibitem{S7}
P.~Moetakef, T.~A.~Cain, D.~G~Ouellette, J.~Y~Zhang, D.~O.~Klenov,
A.~Janotti, Ch.~G.~Van de Walle, S.~Rajan, S.~J.~Allen, and
S.~Stemmer, Appl.~Phys.~Lett. {\bf 99}, 232116 (2011).

\bibitem{S8}
C.~A.~Jackson and S.~Stemmer, Phys.~Rev.~B  {\bf 88}, 180403 (2013).

\bibitem{S9}
J.~Biscaras, N.~Bergeal, A.~Kushwaha, T.~Wolf, A.~Rastogi,
R.\,C.~Budhani, and J.~Lesueur, Nature Communications \textbf{1}, 89
(2010).

\bibitem{S12}
N.~Nakagawa, H.\,Y.~Hwang, and D.\,A.~Muller, Nature Maner.
\textbf{5}, 204 (2006).

\bibitem{S14}
P.~Hohenberg and W.~Kohn,
Phys.~Rev.~{\bf 136}, B864 (1964).

\bibitem{S17}
J.~P.~Perdew, K.~Burke, and M.~Ernzerhof, Phys.~Rev.~Lett.~{\bf 77},
3865 (1996).


\bibitem{S18}
MedeA\textsuperscript{\textregistered}-2.20, Materials Design, Inc.,
San Diego, CA, USA (2015).

\bibitem{S15} G.~Kresse, and J.~Furthm\"uller,
Comp.~Mat.~Sci.~ {\bf 6}, 15 (1996).


\bibitem{S16}  S.\,L.~Dudarev, G.\,A.~Botton, S.\,Y.~Savrasov, C.\,J.~Humphreys, and A.\,P.~Sutton,
Phys.~Rev.~B~{\bf 57}, 1505 (1998).

\bibitem{piyanzina}
I.~I.~Piyanzina, T.~Kopp, Yu.~V.~Lysogosrkiy, D.~Tayurskii, and
V.~Eyert, J.~Phys.: Condens.~Matter {\bf 29}, 095501 (2017).


\bibitem{LCO_data}
M.\,T.\,Czy\ifmmode \dot{z}\else \.{z}\fi{}yk,  and
G.\,A.\,Sawatzky,  Phys. Rev. B \textbf{49}, 14211 (1994).

\bibitem{svane}
A.\,Svane, Phys.~Rev.~let, \textbf{68}, 1900 (1992).

\bibitem{exp_BTO}
C.\,J.\,Xiao, , C.\,Q.\,Jin,  X.\,H.\,Wang, Materials chemistry and physics, \textbf{111}, 209 (2008).

\bibitem{exp_BTO2}
S.\,H.\,Wemple, Phys.~Rev.~B \textbf{2}, 2679 (1970).


\bibitem{S30}
V.~J.~Emery, and S.\,A.~Kivelson, Nature~{\bf 374}, 434 (1995).

\bibitem{S31}
A.\,F.~Bangura, J.\,D.~Fletcher, A.~Carrington, J.~Levallois,
M.~Nardone,B.~Vignolle, P.\,J.~Heard, N.~Doiron-Leyraud, D.~LeBoeuf,
L.~Taillefer, S.~Adachi, C.~Proust, and N.~E. Hussey,
Phys.~Rev.~Lett. {\bf 100}, 047004 (2008).


\bibitem{brandt}
E.\,H.\,Brandt and M.\,Indenbom, Phys. Rev. B \textbf{48}, 12893
(1993).

\bibitem{Mul}
P.-G. de Gennes, Physical Review \textbf{118}, 141 (1960).

\bibitem{LaSr}
Elbio Dagotto, Takashi Hotta, Adriana Moreo, Physics Reports
\textbf{344}, Issues 1-3, 1-153 (2001).


\bibitem{S62}
D. P.Pavlov, I. I. Piyanzina, V. I. Muhortov, A. I. Balbashov, D.
A.Tauyrskii, I. A. Garifullin, R. F. Mamin, JETP Letters
\textbf{106}, 440-444 (2017).


\bibitem{SS7}
Y. Wang, M. K. Niranjan, J. D. Burton, J. M. An, K. D. Belashchenko,
and E.Y. Tsymbal, Phys. Rev. B 79, 212408 (2009).


\bibitem{Bozovic}
I. Bozovic, G. Logvenov, M. A. J. Verhoeven, P. Caputo, E. Goldobin,
and M. R. Beasley, Phys. Rev. Lett. \textbf{93}, 157002 (2004).


\bibitem{S61}
Xiang Leng, Javier Garcia-Barriocanal, Shameek Bose, Yeonbae Lee,
and A. M. Goldman, Phys.~Rev.~Lett. \textbf{107}, 027001 (2011).


\end{thebibliography}
\end{document}